\documentclass[prb,twocolumn,preprintnumbers,amsmath,amssymb]{revtex4}
\usepackage[dvips]{graphicx}
\usepackage{dcolumn}
\usepackage{bm}
\usepackage{times}
\usepackage{mathptmx}
\usepackage{color}

\begin{document}

\title{Observation of Coherent Helimagnons and Gilbert damping in an Itinerant Magnet}

\author{J. D. Koralek$^{1,*,\text\textdagger}$, D. Meier$^{2,*,\text\textdagger}$, J. P. Hinton$^{1,2}$, A. Bauer$^{3}$, S. A. Parameswaran$^{2}$, A. Vishwanath$^{2}$, R. Ramesh$^{1,2}$ R. W. Schoenlein$^{1}$, C. Pfleiderer$^{3}$, J. Orenstein$^{1,2}$}

 \affiliation{$^1$Materials Science Division, Lawrence Berkeley National Laboratory, Berkeley, California 94720, USA}
 \affiliation{$^2$Department of Physics, University of California, Berkeley, California 94720, USA}
 \affiliation{$^3$Physik Department E21, Technische Universit\"at M\"unchen, D-85748 Garching, Germany}
\begin{abstract}
We study the magnetic excitations of itinerant helimagnets by applying time-resolved optical spectroscopy to Fe$_{0.8}$Co$_{0.2}$Si. Optically excited oscillations of the magnetization in the helical state are found to disperse to lower frequency as the applied magnetic field is increased; the fingerprint of collective modes unique to helimagnets, known as helimagnons. The use of time-resolved spectroscopy allows us to address the fundamental magnetic relaxation processes by directly measuring the Gilbert damping, revealing the versatility of spin dynamics in chiral magnets.
\end{abstract}

\date{Dated
\today}

\maketitle

The concept of chirality pervades all of science, having profound implications in physics, chemistry and biology alike. In solids, relativistic spin-orbit coupling can give rise to the Dzyaloshinskii-Moriya (DM) interaction,\cite{Dzyaloshinskii57a,Moriya60a} imparting a tendency for the electron spins to form helical textures with a well-defined handedness in crystals lacking inversion symmetry. Helical spin order is especially interesting when the magnetism arises from the same electrons responsible for conduction as is the case in doped FeSi which displays unconventional magnetoresitence,\cite{Manyala00a,Manyala04a} helimagnetism,\cite{Beille83a} and the recently discovered Skyrmion lattice.\cite{Muhlbauer09a,Munzer10a} The excitations of helimagnets have been studied over the past 30 years, culminating recently in a comprehensive theory of spin excitations called helimagnons.\cite{Kataoka87a,Belitz06a} Signatures of helimagnons have been observed in neutron scattering\cite{Janoschek10a} and microwave absorption,\cite{Onose12a} yet little is known about their magnetodynamics and relaxation phenomena on the sub-picosecond timescales on which magnetic interactions occur. Understanding the dynamics, however, is of great importance regarding spin transfer torque effects in chiral magnets, and related proposed spintronics applications.\cite{Everschor12a,Jonietz10a,Schulz12a}

In this work we study the dynamics of collective spin excitations in the itinerant helimagnet Fe$_{0.8}$Co$_{0.2}$Si. Our optical pump-probe measurements identify anomalous modes at zero wavevector ($q=0$) which we identify unmistakably as helimagnons. These helimagnons manifest as strongly damped magnetization oscillations that follow a characteristic scaling relation with respect to temperature and magnetic field. The sub-picosecond time resolution of our technique enables determination of the intrinsic Gilbert damping parameter which is found to be one order of magnitude larger than in localized systems, revealing the versatility of the spin-lattice interactions available in the emergent class of DM-driven helimagnets.

Despite being a non-magnetic insulator, FeSi is transformed into an itinerant magnet upon doping with cobalt.\cite{Manyala00a,Aeppli92a} We have chosen Fe$_{0.8}$Co$_{0.2}$Si for our study because it can easily be prepared in high quality single crystals\cite{Neubauer11a} with a reasonably high magnetic ordering temperature $T_{\rm N}$, and its exotic equilibrium properties are well characterized, opening the door for non-equilibrium dynamical studies. Small-angle neutron scattering\cite{Munzer10a} was used to determine the phase diagram and has revealed helimagnetic spin textures below $T_{\rm N} = 30$~K that emerge from the interplay between the ferromagnetic exchange and DM interactions. In zero magnetic field the spins form a proper helix with a spatial period of $\approx 350$~\AA,\cite{Grigoriev09a} whereas finite fields cant the spins along the helix wavevector, $\mathbf k_h$, (see Fig. 1(c)) inducing a conical state with a net magnetization. Sufficiently high fields, $H \geq H_c$, suppress the conical order in favor of field alignment of all spins.  In the experiments reported here, femtosecond pulses of linearly polarized 1.5~eV photons from a Ti:Sapphire oscillator were used to excite a (100) oriented single crystal of Fe$_{0.8}$Co$_{0.2}$Si at near normal incidence. The changes induced in the sample by the pump pulse were probed by monitoring the reflection and Kerr rotation of time-delayed probe pulses from the same laser. In order to minimize laser heating of the sample the laser repetition rate was reduced to 20~MHz with an electro-optic pulse picker, ensuring that thermal equilibrium was reached between successive pump pulses.  Signal to noise was improved by modulation of the pump beam at 100~KHz and synchronous lock-in detection of the reflected probe. Kerr rotation was measured using a Wollaston prism and balanced photodiode. All temperature and field scans presented in this work were performed from low to high $T$ and $\mathbf H||(100)$ after zero-field cooling.

\begin{figure*}[t]
\begin{center}
\includegraphics[width=13.8cm]{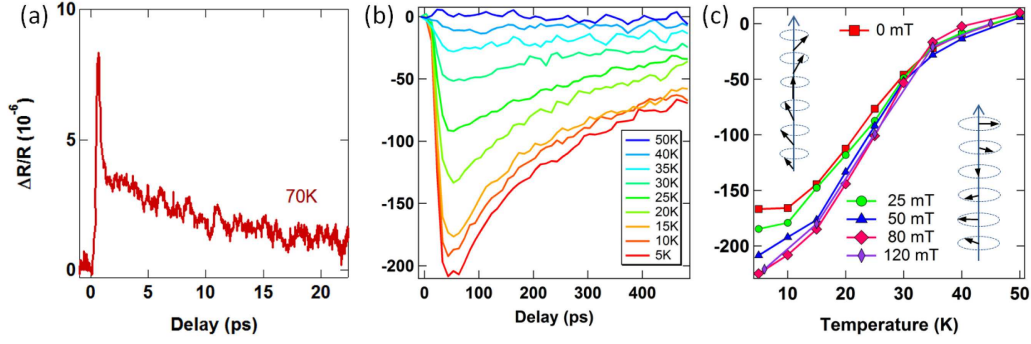}
\end{center}
\caption{\label{fig:1} Time dependence of the pump-induced transient reflectivity $\Delta R/R$ in the (a) paramagnetic and (b) helimagnetic states. The temperature dependence of the maximum $\Delta R/R$ is plotted in (c) for several applied magnetic fields.}
\end{figure*}
 
Fig. 1 shows the transient reflectivity, $\Delta R/R$, as a function of temperature and magnetic field. At high temperature we observe a typical bolometric response from transient heating of the sample by the pump pulse (Fig 1 (a)).\cite{Schoenlein87a} This is characterized by a rapid increase in reflectivity, followed by two-component decay on the fs and ps timescales, corresponding to the thermalization times between different degrees of freedom (electron, spin, lattice, etc.).\cite{Anisimov75a} As the sample is cooled below $T_{\rm N}$, the small thermal signal is beset by a much larger negative reflectivity transient (Fig. 1 (b)) with a decay time of roughly $\tau_R \approx 175$~ps at low temperature (Fig. 3 (b)). A natural explanation for this is that the pump pulse weakens the magnetic order below $T_{\rm N}$, which in turn causes a change in reflectivity via the resulting shift of spectral weight to low energy.\cite{Mena06a} The temperature dependence of the peak $\Delta R/R$ values is plotted in Fig. 1 (c) for several applied fields, showing only weak field dependence.
 
To access the magnetization dynamics more directly we analyze the polarization state of the probe pulses, which rotates by an angle $\theta_K$ upon reflection from the sample surface, in proportion to the component of the magnetization along the light trajectory. The change in Kerr rotation induced by the optical pump, $\Delta\theta_K$, is shown in Fig. 2 as a function of temperature and field. The upper panels show temperature scans at fixed magnetic field, while a field scan at fixed temperature is shown in panel (d). We observe that $\Delta\theta_K$ changes sign as $\mathbf H$ is reversed (not shown), and goes to zero as $\mathbf H$ goes to zero or as temperature is raised above $T_{\rm N}$. Oscillations of the magnetization are clearly visible in the raw data below 25~K in the helimagnetic phase.

\begin{figure*}[t]
\begin{center}
\includegraphics[width=10.8cm]{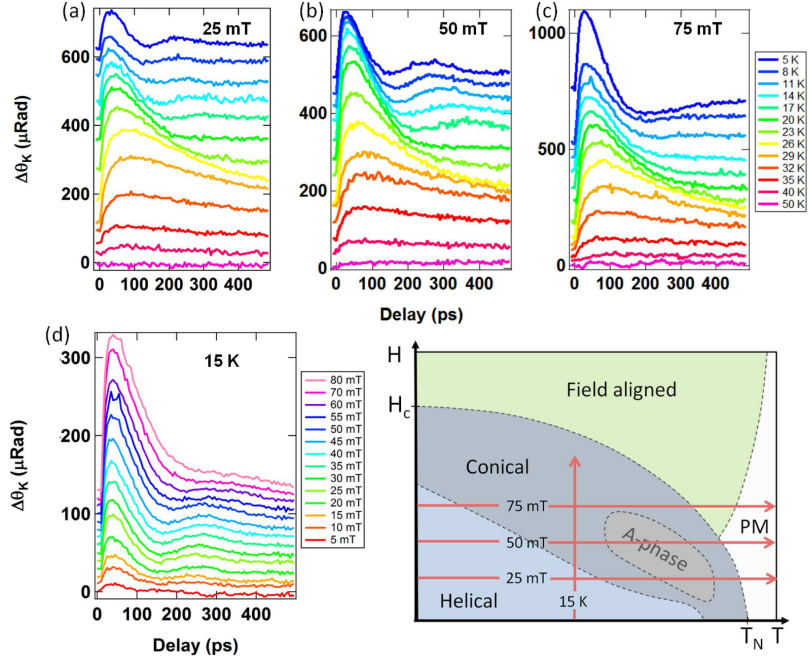}
\end{center}
\caption{\label{fig:2} (a),(b),(c) Time dependence of the pump-induced change in Kerr rotation, $\Delta\theta_K$, as a function of temperature for several applied magnetic fields. (d) $\Delta\theta_K$ as a function of magnetic field for several temperatures. Curves are offset for clarity. Also shown is a schematic phase diagram, adapted from Reference~\onlinecite{Munzer10a}, with red arrows illustrating the temperature and field scans used in (a)-(d).}
\end{figure*}

In order to analyze the magnetization dynamics, we use a simple phenomenological function that separates the oscillatory and non-oscillatory components seen in the data.  It consists of a decaying sinusoidal oscillation,
\begin{equation}
 \Delta\theta_K=e^{-\frac{t}{\tau_K}}\left[A+B\sin (\omega t)\right]
\end{equation}
with a time dependent frequency,
\begin{equation}
\omega(t)=2\pi f_0 \left[ 1+0.8\left(e^{-\frac{t}{\tau_K}}\right)\right]
\end{equation}
which decays to a final value $\omega_0$. We emphasize that there is only a single decay time $\tau_K$ describing the magneto dynamics, and it is directly related to the Gilbert damping parameter $\alpha=(2\pi f_0 \tau_K)^{-1}$.  This function produces excellent fits to the data as illustrated in Fig. 3 (a), allowing accurate extraction of the oscillation frequencies and decay times shown in Figs. 3 (b)-(d). The oscillation frequency is reduced as either field or temperature is increased, while the decay time $\tau_K$ is roughly constant and equal to $\tau_R$ below 20~K. As the temperature is raised towards the phase transition, the relaxation time $\tau_K$ diverges, which can be understood in terms of a diverging magnetic correlation length due to the presence of a critical point. The similarity between the decay times $\tau_R$ and $\tau_K$ within the ordered phase reflects strongly correlated charge and spin degrees of freedom, and supports the notion that $\Delta R/R$ is determined by the magnetic order.

\begin{figure*}[t]
\begin{center}
\includegraphics[width=9.5cm]{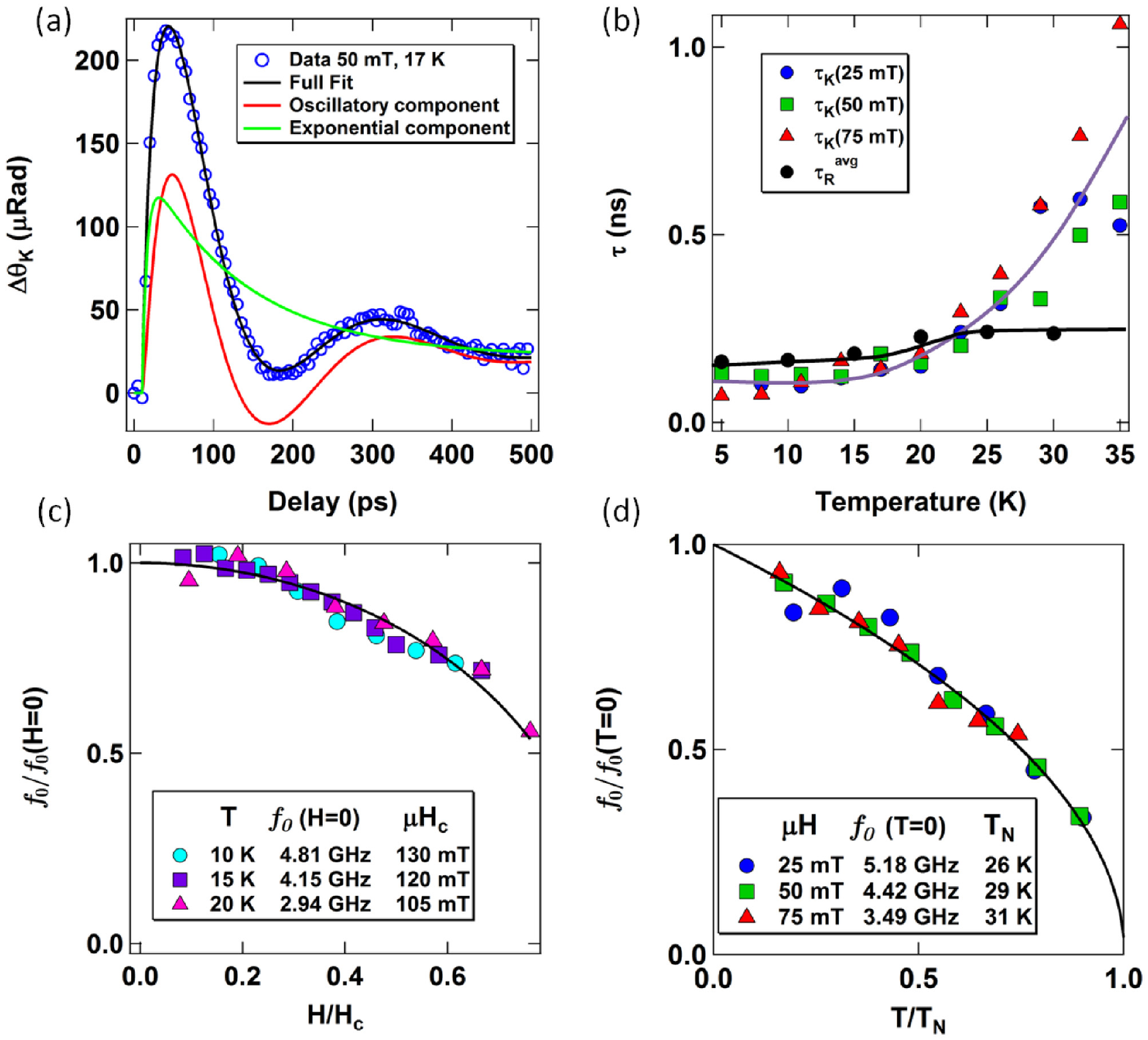}
\end{center}
\caption{\label{fig:3} (a) Exemplary $\Delta\theta_K$ oscillation data (blue circles) and fit (black line) using the model described in the text.  The fit is decomposed into an exponential term (green curve) and an oscillatory term (red curve). The fitting function uses a single time constant $\tau_K$ for all terms which is plotted in panel (b) as a function of temperature and field. For comparison we also plot the decay time of the reflectivity, $\tau_R$, averaged over all fields. The solid lines are guides to the eye. Panels (c) and (d) show the reduced magnetization oscillation frequency for field scans and temperature scans respectively, and solid lines are fits to the data as described in the main text.}
\end{figure*}

The magnetic oscillation frequency reaches $f_0\approx 4.8$ GHz at low temperature, which corresponds to a Larmor precession of spins subjected to a field of 170~mT, which is roughly the critical field $H_c$ required to destroy the spin helix. This, together with the fact that the oscillation frequency is nonzero only in the helical state, suggests that the oscillations are coming from excitations unique to the helical structure. It is well known that magnetization oscillations can be optically induced by ultrafast generation of coherent magnons,\cite{Kampen02a,Kalashnikova08a,Talbayev08a} however, ordinary magnons cannot explain our data as their frequency would increase with $\mathbf H$, opposite to what is seen in Fig. 3(c).  

Based on these observations, we propose the following interpretation of our results: In the helical magnetic phase, the pump photons weaken the magnetic order through the ultrafast demagnetization process.\cite{Kirilyuk10a} As described above, this reduction in magnetic order gives rise to a decrease in the reflectivity at 1.5~eV which is nearly field independent. As a magnetic field is applied the spins become canted along the helix wavevector, giving rise to a macroscopic magnetization which we observe in Kerr rotation via its component along the probe light trajectory. The demagnetization from the pump is responsible for the initial peak seen in the $\Delta\theta_K$ time traces, and is captured by the exponential component of our fitting function (green curve in Fig 3 (a)). The pump photons also launch a coherent spin wave, giving rise to the oscillations in $\Delta\theta_K$ (red curve in Fig. 3 (a)). The form of the oscillatory component goes like $\sin (\omega t)$ rather than $[1-\cos (\omega t)]$, suggesting impulsive stimulated Raman scattering as the mechanism of excitation.\cite{Kalashnikova08a} The anomalous field dependence shown in Fig. 3 (c) leads to the unambiguous conclusion that the optically excited spin waves are the fundamental modes of helimagnets termed helimagnons.\cite{Belitz06a} Specifically, the optically accessible helimagnon mode consists of the constituents of the spin helix precessing in-phase about their local effective field. Since this local effective field is reduced during the ultrafast demagnetization process, the oscillation frequency decreases as a function of time delay as the field recovers, necessitating the time dependent frequency in Eq. 1. The ability to resolve helimagnons with femtosecond time resolution at $q=0$ is unique to our optical probe, and compliments neutron scattering which is restricted to mapping helimagnon bands at higher $q$. This region of reciprocal space is particularly interesting in the case of helimagnets as the periodicity introduced by the helical spin texture generates bands that are centered at $q=\pm k_h$ and therefore have finite frequency modes at $q=0$ even in the absence of a gap. This is in contrast to ordinary magnons in which the bands are generally centered at $q=0$ so that the associated mode has zero frequency. We note that our observations are in agreement with previous work on the collective modes of skyrmions\cite{Mochizuk12a} which coexist with helimagnons in the A-Phase (see Fig. 3).\cite{Onose12a} The appearance of these modes is not expected in our data as their corresponding oscillation periods exceed the observed damping time in Fe$_{0.8}$Co$_{0.2}$Si.

In order to quantitatively test the helimagnon interpretation we take the expression for the $q=0$ helimagnon frequency in an external magnetic field,
\begin{equation}
f_0=g \mu_{\rm B}H_c\sqrt{1+\cos ^2 \theta}
\end{equation}
where $g$ is the effective electron $g$-factor, $\mu_{\rm B}$ is the Bohr magneton, and $\frac{\pi}{2}-\theta$ is the conical angle i.e. the amount the spins are canted away from $\mathbf k_h$ by the applied field $\mathbf H$. Ignoring demagnetization effects of the spin waves themselves, we can write $\sin \theta=\frac{H}{H_c}$, where $H_c$ is the critical field at which the spins all align with the field and the helimagnon ceases to exist as a well-defined mode. Then we obtain,\cite{Kataoka87a}
\begin{equation}
f_0=g \mu_{\rm B}H_c\sqrt{1-\frac{1}{2}\left(\frac{H}{H_c}\right)^2}
\end{equation}
which expresses the magnon frequency as a function of applied field. This expression fits the data remarkably well as shown in Fig. 3 (c), capturing the decrease in frequency with increasing $H$ which is unique to helimagnons. However, due to the fact that the oscillation period exceeds the damping time for fields above 75~mT, it is not possible to extract the value of the critical field $H_c$ in this system. The solid line in Fig. 3 (d) is a fit to the form $f_0\propto\sqrt{1-\frac{T}{T_{\rm N}}}$ which gives $T_{\rm N}$ as a function of $H$ in reasonable agreement with published data.\cite{Munzer10a}

The Gilbert damping parameter can be directly obtained from the measured decay times through the relation $\alpha=(2\pi f_0 \tau_K)^{-1}$, which gives a value of $\alpha\approx 0.4$ for the helimagnetic phase of Fe$_{0.8}$Co$_{0.2}$Si. This is an order of magnitude larger than what was seen in insulating Cu$_2$OSeO$_3$,\cite{Onose12a} where helimagnetism arises from localized rather than itinerant spins.  The contrast in dynamics between these systems is critical in the context of potential spintronic applications based on helimagnetism where there is a tradeoff between fast switching which requires large damping, and stability which relies on low damping.

In summary, this work demonstrates ultrafast coherent optical excitation of spin waves in an itinerant DM-driven spin system and reveals the underlying spin dynamics.  We identify these excitations as helimagnons through their anomalous field dependence and explain our observations with a comprehensive model. Our experiments directly yield the intrinsic Gilbert damping parameter, revealing a striking difference in spin relaxation phenomena between itinerant and localized helimagnets.  The results elucidate the dynamics of collective modes common to the actively studied B20 transition metal compounds that codetermine their performance in potential spin based applications.

Acknowledgments: The work in Berkeley was supported by the Director, Office of Science, Office of Basic Energy Sciences, Materials Sciences and Engineering Division, of the U.S. Department of Energy under Contract No. DE-AC02-05CH11231. C.P. and A.B. acknowledge support through DFG TRR80 (From Electronic Correlations to Functionality), DFG FOR960 (Quantum Phase Transitions), and ERC AdG (291079, TOPFIT). A.B. acknowledges financial support through the TUM graduate school. D.M. acknowledges support from the Alexander von Humboldt foundation and S.A.P. acknowledges support from the Simons Foundation. C.P. and A.B. also thank S. Mayr, W. M\"unzer, and A. Neubauer for assistance.\\

$^{*}$ These authors contributed equally to this work.


\end{document}